\documentclass[a4paper,11pt]{article}
\usepackage{pos}

\title{The four-gluon and ghost-gluon vertices in the Landau gauge from lattice simulations}

\author[a,b]{Nuno Brito}
\author[b]{Manuel Cola\c{c}o}
\author[b]{Orlando Oliveira}
\author*[b]{Paulo J. Silva}

\affiliation[a]{Centre for Mathematical Sciences, University of Plymouth, UK}

\affiliation[b]{CFisUC, Department of Physics, University of Coimbra, Portugal}

\emailAdd{nmrbrito2000@gmail.com}
\emailAdd{manuel.sc.colaco@gmail.com}
\emailAdd{orlando@uc.pt}
\emailAdd{psilva@uc.pt}

\abstract{The computation of the four-gluon and ghost-gluon vertices in the Landau gauge using
high statistical lattice ensembles for $32^4$ and $48^4$ volumes is addressed. For the four-gluon 
vertex, our previous results for the collinear kinematics are updated allowing to get a better 
coverage of the IR region. Furthermore, the one-particle irreducible ghost-gluon Green function 
in the soft gluon limit is computed covering, with precision, a large momentum region.}

\FullConference{The XVIth Quark Confinement and the Hadron Spectrum Conference (QCHSC24)\\
 19-24 August, 2024\\
 Cairns Convention Centre, Cairns, Queensland, Australia\\}

\begin{document}
\maketitle

\section{The Four-Gluon Vertex}

The four-gluon one-particle irreducible Green function (1PI) is a complex object whose tensor decomposition requires a large basis of operators. However, in
lattice simulations only the full Green functions $\mathcal{G}$, that combines 1PI Green functions and propagators,
are measured. The evaluation of the 1PI Green function requires, necessarily, performing some kind of projection to be able to access the form factors. An example of such a situation appears in the calculation of the ghost-gluon vertex; see Sec. \ref{Sec:ghost-gluon} below. 

The diagrammatic representation of the four-gluon full Green function is

\begin{figure}[h]
      \centering
	  \includegraphics[width=12cm]{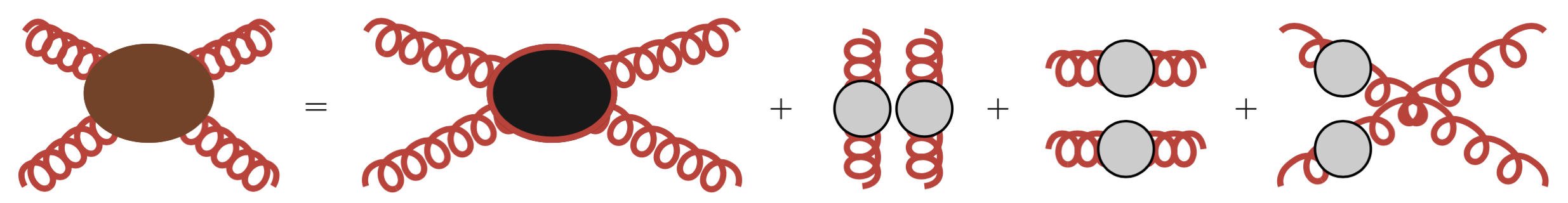} 
\end{figure}
\noindent
where the grey blobs represent 1PI Green functions, and the first term in the r.h.s. stands for the contribution of the connected diagrams. This last Green function
is represented in terms of 1PI functions as

\begin{figure}[h]
      \centering
	  \includegraphics[width=12cm]{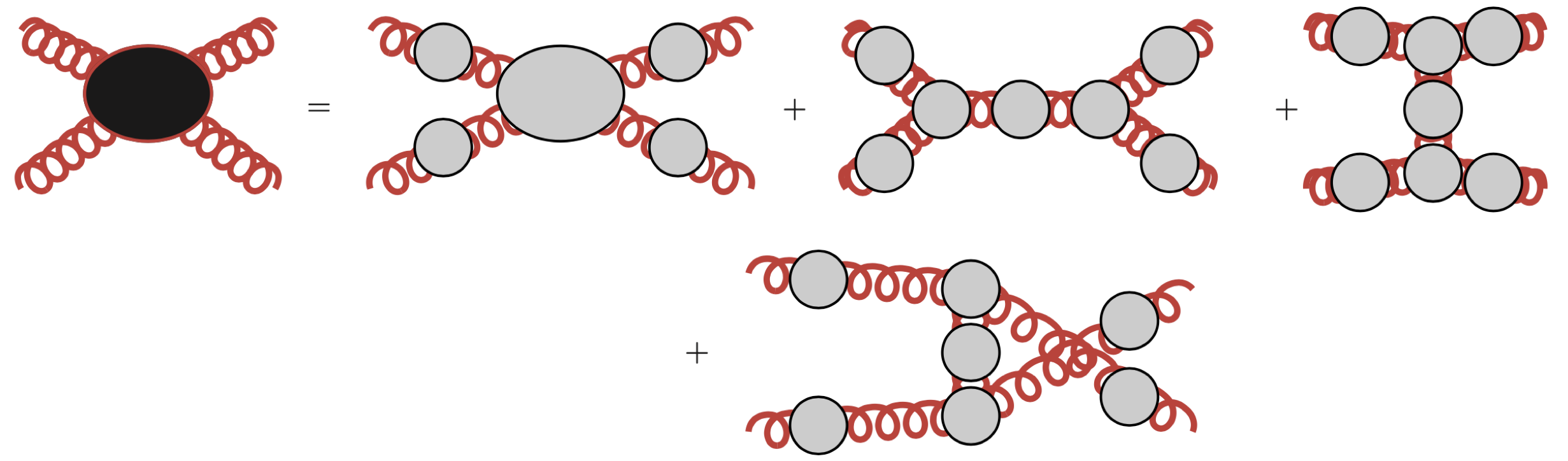}  
\end{figure}
\noindent
and, therefore, the measurement of the four-gluon 1PI function contribution, that  appears only in the first diagram in the r.h.s. of the above decomposition,
requires the subtraction of many contributions of different nature and are associated either to disconnected diagrams or include the contribution of
the three-gluon 1PI Green function. However, when the external momenta are proportional to each other,  i.e. when $p_i \propto p_j$ and $p_i \ne p_j$,
only the diagram with the four-gluon 1PI Green function contributes to the full Green function, see the discussion in \cite{Colaco24} where further details can be found. 
Moreover, the number of tensors that define a basis for the four-point 1PI function is greatly reduced, specially in the Landau gauge due to the orthogonality of the gluon propagator.
For this kinematics the total number of independent tensors is reduced to three and the form factors to be measured are associated with the operators
\begin{eqnarray}
 \widetilde{\Gamma}^{(0)} \,  ^{abcd}_{\mu\nu\eta\zeta}  & =  &
   f_{abr} f_{cdr} \left( g_{\mu\eta} g_{\nu \zeta} - g_{\mu\zeta} g_{\nu \eta} \right) +
               f_{acr} f_{bdr} \left( g_{\mu\nu} g_{\eta\zeta} - g_{\mu\zeta} g_{\nu \eta}   \right)  +
               \nonumber \\  & & \qquad\qquad
               f_{adr} f_{bcr} \left( g_{\mu\nu} g_{\eta\zeta}  - g_{\mu\eta} g_{\nu \zeta}  \right) \nonumber\\
 \widetilde{\Gamma}^{(1)} \,  ^{abcd}_{\mu\nu\eta\zeta}  & = &
           d_{abr} d_{cdr} \left( g_{\mu\eta} g_{\nu \zeta} + g_{\mu\zeta} g_{\nu \eta} \right) +
           d_{acr} d_{bdr} \left( g_{\mu\zeta} g_{\nu \eta} + g_{\mu\nu} g_{\eta\zeta} \right)  +
           \nonumber \\  & & \qquad\qquad
          d_{adr} d_{bcr} \left( g_{\mu\nu} g_{\eta\zeta}  + g_{\mu\eta} g_{\nu \zeta}  \right) \nonumber\\
 \widetilde{\Gamma}^{(2)} \,  ^{abcd}_{\mu\nu\eta\zeta}   & = &
               \Big(  \delta^{ab}  \, \delta^{cd} + \delta^{ac}  \, \delta^{bd} + \delta^{ad}  \, \delta^{bc}  \Big) 
               \Big(  g_{\mu\nu}  \, g_{\eta\zeta} + g_{\mu\eta}  \, g_{\nu\zeta} + g_{\mu\zeta}  \, g_{\nu\eta} \Big) 
\end{eqnarray}
that define a basis for the collinear kinematics. The associated form factors $F^{(i)}$ are measured from the  amputated Green function.
The tensor $\widetilde{\Gamma}^{(0)}$ is the same tensor structure that define the tree level perturbative Feynman rule. 
Further details can be found in \cite{Colaco24}. The lattice results therein were updated in \cite{Oliveira:2025tfq}. See also \cite{Sil04} for definitions and notation. 

The results reported here use  pure Yang-Mills configuration ensembles, in the Landau gauge, generated with the Wilson
action with a $\beta = 6.0$ (inverse lattice spacing of 1.943 GeV) -- see \cite{Duar16} and references therein. In the following results for a $32^4$ lattice (9038 configurations) and a
$48^4$ (9035 configurations) will be reported. Due to Bose symmetry the amputated form factors are functions of
the combination of momenta $s = \sum_i p^2_i / 4$. 

\begin{figure}[h]
      \centering
      \includegraphics[width=7.8cm]{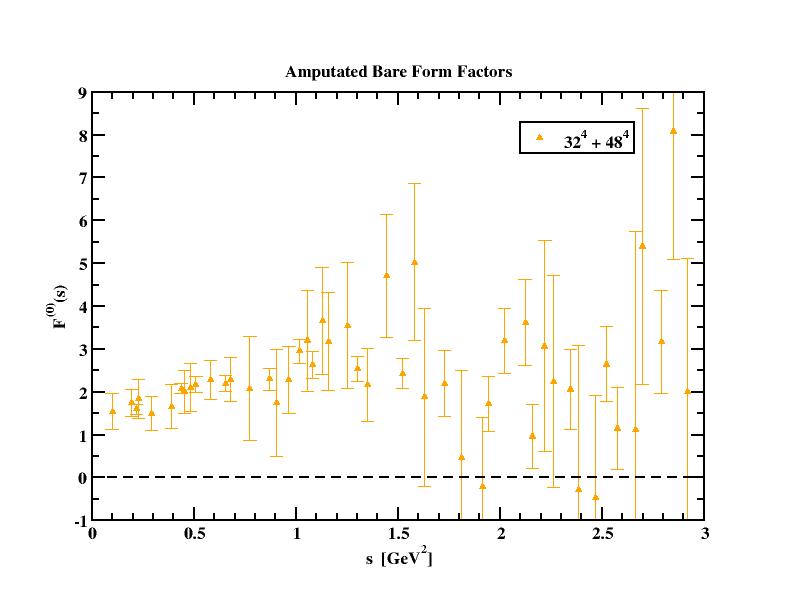}  \hspace*{-7mm}
%
	  \includegraphics[width=7.8cm]{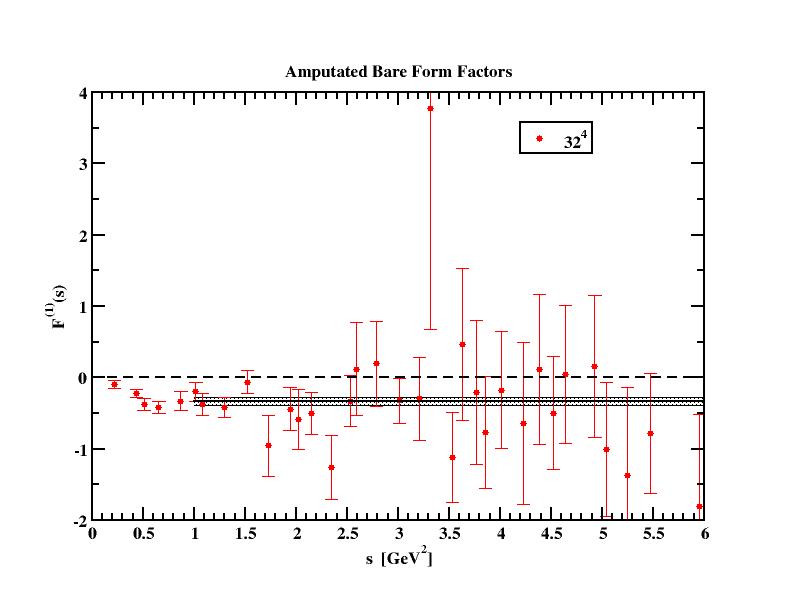}  
%
	  \includegraphics[width=7.8cm]{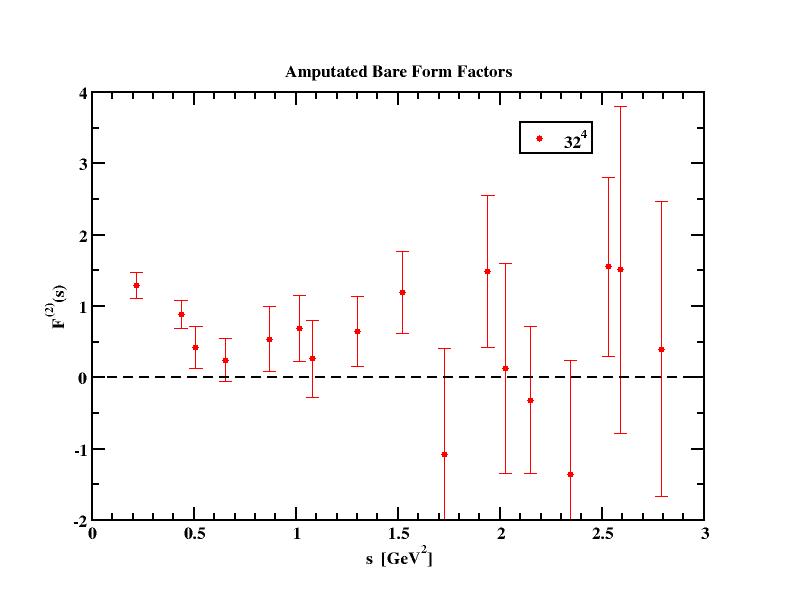}  
\caption{Amputated form factors that describe the four-gluon 1PI Green function. \label{Fig1}}
\end{figure}

In general, the results of the simulations for $F^{(i)}$, see Fig. \ref{Fig1},

\begin{itemize}
\item have a good signal-to-noise ratio for momenta up to $s = \sum_i p^2_i / 4 \sim 1.5$ GeV$^2$. For large momentum the
division by (small) values of the propagators result in large statistical errors;

\item  it is observed a hierarchy between the various form factors with $F^{(0)}$ and $F^{(2)}$ being much larger than $F^{(1)}$;

\item the form factor $F^{(0)}$ seems to be essentially constant, while the data suggest that $F^{(2)}$ grow in the infrared region as $s$ approaches
zero.
\end{itemize}

\vspace{-0.2cm}


\newpage
\section{The ghost-gluon vertex \label{Sec:ghost-gluon}}

The ghost-gluon 1PI vertex

\vspace{-0.2cm}
\begin{figure}[h]
      \centering
	  \includegraphics[width=4cm]{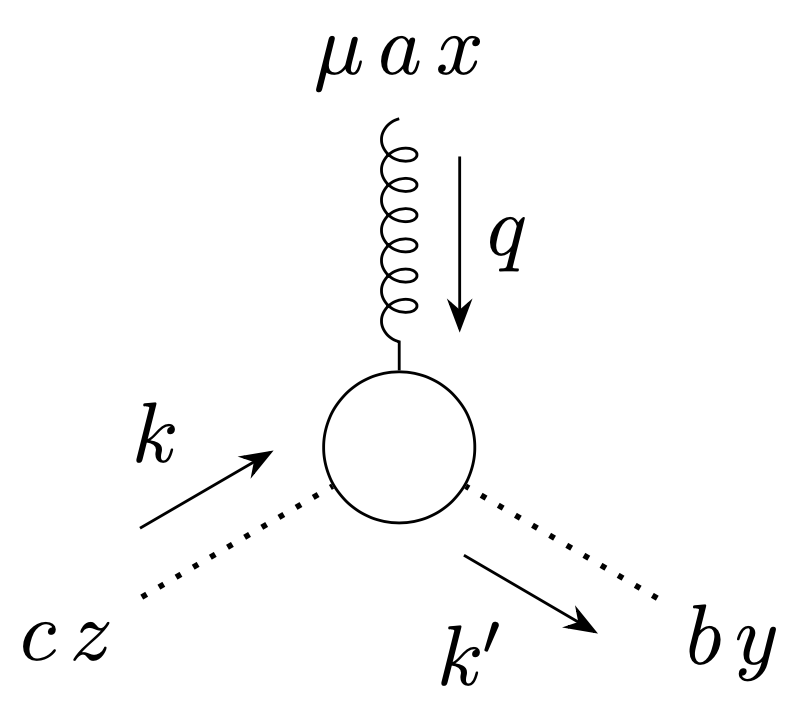} 
\end{figure}

\noindent
is described by
\begin{displaymath}
    \Gamma^\mu = i \, g \, f_{abc} \bigg( k^\prime_\mu \, H_1(k^2, \, k^{\prime  ^2}, \, q^2 ) + q_\mu \, H_2(k^2, \, k^{\prime  ^2}, \, q^2 )\bigg)
\end{displaymath}
where $g$ is the strong coupling constant and $H_i$  are Lorentz scalar form factors.
$\Gamma^\mu$ is important for the QCD dynamics, determining, for example, the ghost propagator.
In the Landau gauge, due to the orthogonality of gluon propagator, the contribution of the 
form factor $H_2$ vanishes and only $H_1$ can be accessed.
The tree level lattice 1PI is
\begin{displaymath}
    \Gamma^{Lat}_\mu = i \, g \, f_{abc} \, k^\prime_\mu \,  \cos \left( \frac{a \, k_\mu}{2}  \right)
\end{displaymath}
with
\begin{displaymath}
   k_\mu = \frac{ 2 \, \pi}{a \, N_\mu} \, n_\mu \ , \qquad n_\mu = 0, \dots, N_\mu/2 \ ,
\end{displaymath}
 $a$ being the lattice spacing, $N_\mu$ the number of lattice points in direction $\mu$, and
 $k^\prime_\mu = (2/a) \, \sin ( a \, k_\mu / 2)$ being the improved momenta that reproduces
 the naive lattice momenta up to corrections of $\mathcal{O}(a^2)$. 
 The form factor $H_1$ is given by the Lorentz-color contraction
\begin{displaymath}
   H_1  = \frac{ \Gamma_{(i)} \, \mathcal{G} }{ \Gamma_{(i)} \, D_{gl} \, D_{gh} \, D_{gh} \, \Gamma_{(i)} } 
\end{displaymath}
where all indices and momenta were omitted to simplify the notation. 
$D_{gl}$ stands for the gluon propagator, $D_{gh}$ for the ghost propagator and the vertex
$\Gamma_{(i)}$ represents either $\Gamma_{Lat}$ of the continuum vertex $\Gamma_{Cont}$ (differs
from $\Gamma_{Lat}$  by the $\cos$-term).

\begin{figure}[h]
	\centering
	\includegraphics[scale=0.3]{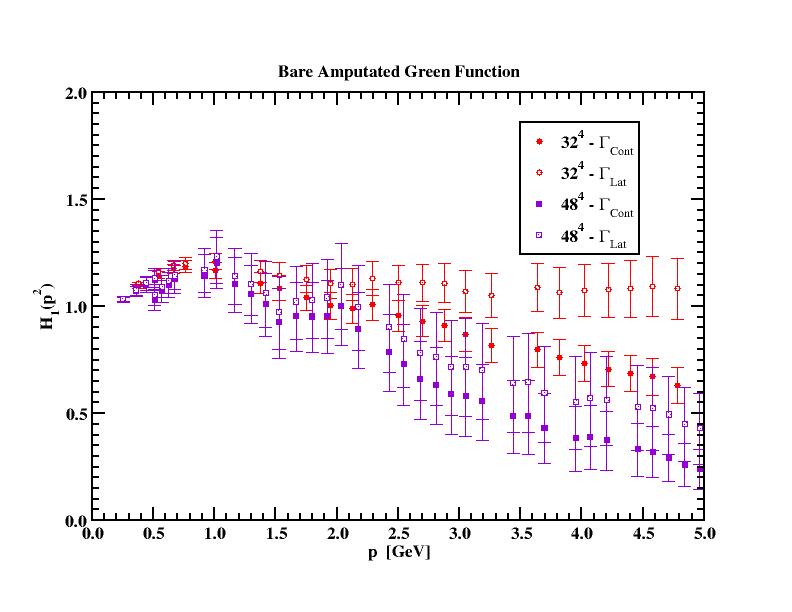}
	\caption{The ghost-gluon vertex data. \label{Fig:ghost-gluon}}
\end{figure}

The results for $H_1$ use a subset of the ensembles mentioned previously (3000 configurations
for $32^4$; 2000 configurations for $48^4$) that were generated for the work in \cite{Colaco24}. 
The lattice data can be seen in Fig. \ref{Fig:ghost-gluon}. See also the Lattice 2024 proceeding \cite{Brito:2024aod}.
Previous lattice calculations of $H_1$ can be found in \cite{Cuc08,Ma20,Ilg07}.

\section{Summary and Conclusions}	

In this proceeding we report on-going lattice calculations of the four-gluon and ghost-gluon vertices using lattice
QCD simulations in the Landau gauge. In both cases, large statistical ensembles are considered but for a unique lattice
spacing, preventing from studying the continuum limit. Previous lattice studies of the gluon propagator suggest
that the finite size effects are small for the type of ensembles considered herein \cite{Oliveira:2012eh}.

In what concerns the four-gluon vertex, although being a difficult lattice calculation, the signal-to-noise is reasonably good below $p < 1.5$ GeV. 
Moreover, the numerical simulations considered here agree, at least at a qualitative level, with recent continuum calculations \cite{Ag24,Bar24} of the same
form factors. Other kinematical configurations were also investigated using both continuum and lattice methods in \cite{Aguilar:2024dlv}.

The ghost-gluon vertex in the soft gluon limit was computed also with subsets of the configurations used to investigate the four-gluon vertex.
Results obtained for the measured form factor using  continuum and lattice versions of the tensor structure were investigated
to check for finite size effects. The numerics show that in both cases, the two results agree over a large range of momenta and specially at IR, suggesting
that finite size effects are under control for momentum up to 3 GeV. For larger momenta, i.e. for  $k \gtrsim 3$ GeV the lattice spacing effects need to
be better understood. Oure result are in good agreement with previous lattice calculations \cite{Ilg07,Ag24,Bar24}, despite their large statistical errors for the SU(3) case.

The determination of the four-gluon and the ghost-gluon vertices need larger gauge ensembles to reduce the statistical errors, together with a better understanding
of the systematics.

\section*{Acknowledgements}

This work was supported by  FCT – Funda\c{c}\~ao para a Ci\^encia e a Tecnologia, I.P., under Projects Nos. UIDB/04564/2020,
 (\url{https://doi.org/10. 54499/UIDB/04564/2020}), 
UIDP/04564/2020.
(\url{https://doi.org/10.54499/UIDP/04564/2020}) and CERN/FIS- PAR/0023/2021. 
N. B. acknowledges travel support provided by UKRI Science and Tecnhology Facilities (ST/X508676/1). 
P. J. S. acknowledges financial support from FCT contract CEECIND/00488/2017. 
(\url{https: //doi.org/10.54499/CEECIND/00488/2017/CP1460/CT0030}). 
The authors acknowledge the Laboratory for Advanced Computing at the University of Coimbra (http://www.uc.pt/lca) for providing access to the HPC resources.
Access to Navigator was partly supported by the FCT Advanced Computing Projects 2021.09759.CPCA, 2022.15892.CPCA.A2, 2023.10947.CPCA.A2.

\end{document}